\documentclass[11pt,preprintnumbers,titlepage,nofootinbib]{revtex4-2}%
\usepackage[latin9]{inputenc}
\usepackage{amsmath}
\usepackage{amssymb}
\usepackage{graphicx}
\usepackage{wasysym}
\usepackage[unicode=true,  bookmarks=false,  breaklinks=false,pdfborder={0 0
1},colorlinks=false]{hyperref}
\usepackage{array}
\usepackage{multirow}
\usepackage{wasysym}
\usepackage{wasysym}
\usepackage{amsfonts}
\usepackage{subfigure}
\usepackage{cleveref}
\usepackage{listings}
\usepackage{xcolor}
\usepackage{array}
\usepackage{url}
\usepackage{color}%
\hypersetup{
colorlinks,linkcolor=blue,citecolor=blue,urlcolor=blue}
\makeatletter

\@ifundefined{textcolor}{}{
\definecolor{BLACK}{gray}{0}
\definecolor{WHITE}{gray}{1}
\definecolor{RED}{rgb}{1,0,0}
\definecolor{GREEN}{rgb}{0,1,0}
\definecolor{BLUE}{rgb}{0,0,1}
\definecolor{CYAN}{cmyk}{1,0,0,0}
\definecolor{MAGENTA}{cmyk}{0,1,0,0}
\definecolor{YELLOW}{cmyk}{0,0,1,0}
}

\makeatother
\begin{document}
\preprint{CTP-SCU/20260001}
\title{Exceptional Points in Quasinormal Spectra of Hairy Black Holes}
\author{Lang Cheng$^{a}$}
\email{langcheng@stu.scu.edu.cn}
\author{Xiaobo Guo$^{b}$}
\email{guobo@ustc.edu.cn}
\author{Yuhan Li$^{a}$}
\email{leeyhan@stu.scu.edu.cn}
\author{Jun Tao$^{a}$}
\email{taojun@scu.edu.cn}
\author{Peng Wang$^{a}$}
\email{pengw@scu.edu.cn}
\affiliation{$^{a}$College of Physics, Sichuan University, Chengdu, 610065, China}
\affiliation{$^{b}$School of Data Science, Shanghai University of Finance and Economics
Zhejiang College, Jinhua, 321015, China}

\begin{abstract}
Exceptional points (EPs) in quasinormal mode (QNM) spectra are non-Hermitian
degeneracies at which both the eigenvalues and eigenfunctions coalesce. In
this paper, we identify an EP in the scalar QNM spectrum of hairy black holes
in the Einstein-Maxwell-scalar theory by scanning the parameter space. We then
investigate its implications for ringdown signals by extracting QNMs from
time-domain waveforms. Our results show that an EP ansatz, which includes a
resonant contribution containing a term linear in time, provides a more robust
description of ringdown at the EP than the standard ansatz based on a
superposition of independent damped modes. In particular, it captures the
resonant contribution associated with spectral coalescence more naturally, and
even when the waveform is fitted with the standard ansatz, the resulting fit
may still exhibit characteristic features of the EP ansatz.

\end{abstract}
\maketitle
\tableofcontents

\section{Introduction}

The ringdown stage of a perturbed black hole (BH) is governed by its
quasinormal modes (QNMs), whose complex frequencies determine the oscillation
rates and damping times of the spacetime response. QNMs therefore play a
central role in BH spectroscopy and provide a direct link between
gravitational-wave (GW) observations and the properties of compact objects
\cite{Kokkotas:1999bd,Nollert:1999ji,Berti:2009kk,Konoplya:2011qq,Berti:2025hly}%
. In the standard description, the ringdown signal is written as a
superposition of exponentially damped modes. A basic issue, however, is how
reliably these modes can be identified and extracted from time-domain
waveforms, especially when several modes contribute at comparable levels or
when the relevant frequencies are nearly degenerate
\cite{Giesler:2019uxc,Baibhav:2023clw,Takahashi:2023tkb}.

Recent developments have highlighted the non-Hermitian character of the QNM
eigenvalue problem. Because QNMs are defined by purely ingoing boundary
conditions at the horizon and purely outgoing boundary conditions at spatial
infinity, the resulting spectral problem is intrinsically non-Hermitian
\cite{El-Ganainy:2018ksn,Bergholtz:2019deh,Ashida:2020dkc}. As a result, the
spectrum can display pronounced sensitivity to perturbations, pseudospectral
broadening, resonance, and other effects familiar from non-Hermitian theory
\cite{Jaramillo:2020tuu,Cheung:2021bol,Jaramillo:2021tmt,Cao:2024oud,Ianniccari:2024ysv,Yang:2024vor,Chen:2024mon,Cai:2025irl,Mai:2025cva}%
. Among these, exceptional points (EPs) have attracted increasing attention in
BH perturbation theory, especially with the advent of high-precision GW
astronomy
\cite{LIGOScientific:2016aoc,LIGOScientific:2017vwq,LIGOScientific:2018dkp,LIGOScientific:2020tif,LIGOScientific:2025hdt}%
. An EP is a spectral degeneracy at which two eigenvalues and their associated
eigenfunctions coalesce
\cite{Heiss:1999alz,Klaiman:2008zz,Nesterov:2008cg,Dietz:2010bvm,Heiss:2012dx,Schnabel:2017tti,Zhong:2018hmw,Kumar:2021lkq,Ryu:2025gnn}%
. Near such a point, the spectrum develops the characteristic square-root
branch structure, which is associated with a variety of phenomena, including
mode switching, hysteresis, and resonant behavior
\cite{Motohashi:2024fwt,Cavalcante:2024swt,Yang:2025dbn,PanossoMacedo:2025xnf}.

EPs generally require a parameter space of dimension greater than one. A
representative example is provided by massive scalar perturbations of Kerr
BHs, where EPs appear in the parameter space spanned by the BH spin and the
scalar mass
\cite{Cavalcante:2024swt,Cavalcante:2024kmy,Cavalcante:2025abr,Chen:2025sbz,Nakamoto:2026lyo}%
. A related construction can also be realized by adding a Gaussian bump to the
curvature effective potential that governs BH perturbations, and this idea has
recently been extended from isolated EPs to continuous exceptional lines
\cite{Yang:2025dbn,Cao:2025afs,Wu:2025wbp}. By contrast, when only a
restricted one-parameter scan is available, an exact EP is generically
replaced by mode repulsion or an avoided crossing, which can be understood as
the one-dimensional trace of a nearby EP in a higher-dimensional parameter
space
\cite{Onozawa:1996ux,Dias:2021yju,Dias:2022oqm,Davey:2023fin,Motohashi:2024fwt,Oshita:2025ibu,Lo:2025njp,Cook:2014cta,Takahashi:2025uwo}%
.

These features matter not only for the organization of the QNM spectrum in the
frequency domain, but also for the description of ringdown signals in the time
domain. Near an EP, the usual expansion in terms of independent exponentially
damped QNMs is modified by a resonant contribution containing both an
exponential factor and a term linear in time \cite{Yang:2025dbn}.
Consequently, the presence of an EP can leave a characteristic imprint on the
waveform and affect the extraction and interpretation of ringdown signals.
Recent studies have also indicated that an EP based description can capture
near resonant BH ringdown more naturally than the standard superposition of
independent damped modes \cite{PanossoMacedo:2025xnf,Kubota:2025hjk}.

Most existing studies of EPs in BH perturbation theory have focused on BHs
satisfying the no-hair theorem, whereas the corresponding problem in hairy BH
backgrounds remains much less explored. This naturally raises the question of
whether similar non-Hermitian structures also arise in BH solutions with
nontrivial hair. In particular, hairy BHs in the Einstein-Maxwell-scalar (EMS)
theory provide a useful setting in which to investigate this issue. The theory
admits asymptotically flat BH solutions with scalar hair and contains two
independent control parameters, allowing for a richer spectral structure
\cite{Herdeiro:2015waa,Herdeiro:2018wub,Myung:2018jvi,Fernandes:2019rez,Fernandes:2019kmh,Blazquez-Salcedo:2020nhs,LuisBlazquez-Salcedo:2020rqp,Wang:2020ohb,Guo:2021zed,Guo:2021enm,Belkhadria:2023ooc,Melis:2024kfr}%
. More importantly, recent studies of the scalar QNM spectrum of EMS hairy BHs
have identified two distinct families of modes with different spectral
stability properties: peak modes, which are localized near the potential peak
and are associated with the photon sphere, and off-peak modes, which are
localized away from the potential peak and retain an imprint of the valley
structure arising in the double-peaked effective potential \cite{Wang:2025mxe}%
. An avoided crossing between a peak mode branch and an off-peak mode branch
was also observed, suggesting that an EP may emerge once the full
two-parameter space is explored.

In this paper, we study EPs in the scalar QNM spectrum of hairy BHs in the EMS
model and investigate their implications for ringdown signals. Specifically,
we identify an EP in the QNM spectrum and compare the corresponding
time-domain extraction using the standard QNM ansatz and an EP based ansatz.
The remainder of this paper is organized as follows. In Sec.~\ref{sec:setup},
we briefly review the background hairy BH solutions and the numerical
framework. In Sec.~\ref{sec:QS}, we present the QNM spectrum and discuss the
EP structure. In Sec.~\ref{sec:QE}, we analyze the corresponding ringdown
signal and its extraction. We conclude in Sec.~\ref{sec:Con} with a summary
and discussion. Throughout this work, we employ geometrized units with $G=c=1$.

\section{Setup}

\label{sec:setup}

In this section, we outline the background BH solutions and the framework used
to compute the QNM spectrum of a test scalar field. We also summarize our
numerical strategy, including both the frequency-domain spectral method and
the time-domain waveform extraction.

\subsection{Black Hole Solution}

We study the scalar QNM spectrum of static, spherically symmetric BH
backgrounds in the Einstein-Maxwell-scalar (EMS) theory. The theory contains a
real scalar field $\phi$ that is minimally coupled to the spacetime metric and
nonminimally coupled to the electromagnetic field $A_{\mu}$. The dynamics is
governed by the action \cite{Herdeiro:2018wub}
\begin{equation}
S=\int d^{4}x\,\sqrt{-g}\left[  \,R-2\left(  \partial\phi\right)
^{2}-e^{\alpha\phi^{2}}F^{\mu\nu}F_{\mu\nu}\right]  ,
\end{equation}
where $F_{\mu\nu}=\partial_{\mu}A_{\nu}-\partial_{\nu}A_{\mu}$ is the
electromagnetic field strength tensor. The coupling function $e^{\alpha
\phi^{2}}$ encodes the interaction between $\phi$ and $A_{\mu}$, where
$\alpha$ denotes the coupling constant.

Within this framework, the equations of motion admit static, spherically
symmetric, asymptotically flat BH solutions endowed with nontrivial scalar
hair. Adopting the metric ansatz
\begin{equation}
ds^{2}=-N\left(  r\right)  e^{-2\delta\left(  r\right)  }dt^{2}+\frac{dr^{2}%
}{N\left(  r\right)  }+r^{2}d\Omega^{2},
\end{equation}
together with the electrostatic potential $A_{\mu}dx^{\mu}=\Phi(r)\,dt$, the
equations of motion reduce to a set of four coupled ordinary differential
equations for the functions $N(r)$, $\delta(r)$, $\phi(r)$, and $\Phi(r)$:
\begin{align}
N^{\prime}\left(  r\right)   &  =\frac{1-N\left(  r\right)  }{r}-\frac{Q^{2}%
}{r^{3}e^{\alpha\phi^{2}\left(  r\right)  }}-rN\left(  r\right)  \left[
\phi^{\prime}\left(  r\right)  \right]  ^{2},\nonumber\\
\left[  r^{2}N\left(  r\right)  \phi^{\prime}\left(  r\right)  \right]
^{\prime}  &  =-\frac{\alpha Q^{2}\phi\left(  r\right)  }{r^{2}e^{\alpha
\phi^{2}\left(  r\right)  }}-r^{3}N\left(  r\right)  \left[  \phi^{\prime
}\left(  r\right)  \right]  ^{3},\nonumber\\
\delta^{\prime}\left(  r\right)   &  =-r\left[  \phi^{\prime}\left(  r\right)
\right]  ^{2},\\
\Phi^{\prime}\left(  r\right)   &  =\frac{Q}{r^{2}e^{\alpha\phi^{2}\left(
r\right)  }}e^{-\delta\left(  r\right)  },\nonumber
\end{align}
where $Q$ is the electric charge, and primes denote derivatives with respect
to the areal radius $r$. Regularity at the event horizon and asymptotic
flatness at spatial infinity fix the boundary conditions. Owing to a scaling
symmetry of the field equations, the family of solutions can be parameterized
by the dimensionless pair $(\alpha,q)$, where $q\equiv Q/M$, and $M$ denotes
the ADM mass of the BH. Following \cite{Wang:2025mxe}, we construct the hairy
BH solutions using a spectral method.

The hairy BH solutions considered here are also referred to as scalarized
Reissner-Nordstr\"{o}m (RN) BHs. In the EMS theory, the nonminimal coupling
between the scalar and electromagnetic fields can trigger a tachyonic
instability, so that an originally bald RN BH can dynamically evolve into a
scalarized BH. Studying the dynamical properties of these scalarized BHs
therefore provides a useful window into the mechanism of spontaneous
scalarization
\cite{Zhang:2021nnn,Zhang:2023qtn,Zhang:2024wci,Garcia-Saenz:2024beb,Guo:2024cts,Zhang:2025jlb,Garcia-Saenz:2025rbc,Qin:2026axh}%
. In addition, the scalarized BHs can exhibit multiple light rings in the
equatorial plane \cite{Gan:2021xdl,Guo:2023mda}, leading to distinctive
optical signatures in BH imaging
\cite{Gan:2021pwu,Guo:2022muy,Chen:2022qrw,Chen:2023qic,Chen:2024ilc}.

\subsection{Scalar QNM}

For simplicity, we consider a test, massless, neutral scalar field $\Psi$
propagating on the hairy BH background. In the test field approximation,
$\Psi$ is assumed to be sufficiently weak that its backreaction on the
background spacetime can be neglected. The scalar field $\Psi$ obeys the
Klein-Gordon equation
\begin{equation}
\frac{1}{\sqrt{-g}}\partial_{\mu}\!\left(  g^{\mu\nu}\sqrt{-g}\,\partial_{\nu
}\Psi\right)  =0. \label{eq:KG}%
\end{equation}
Exploiting the spherical symmetry of the background spacetime, we decompose
the scalar field as
\begin{equation}
\Psi\left(  t,r,\theta,\varphi\right)  =\frac{1}{r}\sum_{l,m}\psi_{lm}\left(
t,r\right)  Y_{l,m}\left(  \theta,\varphi\right)  .
\end{equation}
Substituting this decomposition into Eq.~\eqref{eq:KG}, we obtain a
one-dimensional Regge-Wheeler-Zerilli-type wave equation,
\begin{equation}
\left(  -\frac{\partial^{2}}{\partial t^{2}}+\frac{\partial^{2}}{\partial
r^{\ast2}}-V_{\mathrm{eff}}\left(  r\right)  \right)  \psi\left(  t,r\right)
=0, \label{eq:TDME}%
\end{equation}
where the tortoise coordinate $r^{\ast}$ is defined by
\begin{equation}
\frac{dr^{\ast}}{dr}\equiv\frac{e^{\delta(r)}}{N(r)},
\end{equation}
and the effective potential is
\begin{equation}
V_{\mathrm{eff}}\left(  r\right)  =\frac{e^{-2\delta(r)}N(r)}{r^{2}}\left[
l(l+1)+1-N(r)-\frac{Q^{2}}{r^{2}e^{\alpha\phi^{2}(r)}}\right]  .
\end{equation}
Here $l$ is the angular quantum number.

To obtain the QNM spectrum, we work in the frequency domain by Fourier
decomposing
\begin{equation}
\psi\left(  t,r\right)  =\int d\omega\,\widetilde{\psi}\left(  \omega
,r\right)  e^{-i\omega t},
\end{equation}
which yields the frequency-domain master equation
\begin{equation}
\left(  \frac{d^{2}}{dr^{\ast2}}+\omega^{2}-V_{\mathrm{eff}}\left(  r\right)
\right)  \widetilde{\psi}\left(  \omega,r\right)  =0. \label{eq:FDME}%
\end{equation}
The QNM frequencies $\omega$ are determined by imposing purely ingoing
boundary conditions at the event horizon ($r=r_{H}$) and purely outgoing
conditions at spatial infinity ($r\rightarrow\infty$). From the asymptotic
analysis, the wave function behaves as
\begin{align}
\widetilde{\psi}\left(  \omega,r\right)   &  \sim\left(  r-r_{H}\right)
^{-i\omega/\sqrt{f_{H}h_{H}}},\qquad r\rightarrow r_{H},\nonumber\\
\widetilde{\psi}\left(  \omega,r\right)   &  \sim e^{i\omega r}\left(
r/r_{H}\right)  ^{-i\left(  f_{I}+h_{I}\right)  \omega/2},\qquad
r\rightarrow\infty.
\end{align}
The coefficients $f_{H}$, $h_{H}$, $f_{I}$, and $h_{I}$ follow from the
near-horizon and asymptotic expansions of the metric functions,
\begin{align}
N\left(  r\right)  e^{-2\delta\left(  r\right)  }  &  \sim f_{H}\left(
r-r_{H}\right)  ,\qquad N\left(  r\right)  \sim h_{H}\left(  r-r_{H}\right)
,\nonumber\\
N\left(  r\right)  e^{-2\delta\left(  r\right)  }  &  \sim1+f_{I}/r,\qquad
N\left(  r\right)  \sim1+h_{I}/r,
\end{align}
where the second line holds for $r\rightarrow\infty$. For stable QNMs with
$\mathrm{Im}(\omega)<0$, the corresponding solutions diverge at both the
horizon and spatial infinity. To facilitate the numerical computation, we
therefore factor out the known asymptotic behavior by writing
\begin{equation}
\widetilde{\psi}\left(  \omega,r\right)  =e^{i\omega r}\left(  r/r_{H}\right)
^{-i\frac{\left(  f_{I}+h_{I}\right)  \omega}{2}}\left(  1-r_{H}/r\right)
^{-\frac{i\omega}{\sqrt{f_{H}h_{H}}}}u\left(  r\right)  .
\end{equation}
Substituting this ansatz into Eq.~\eqref{eq:FDME} yields an ordinary
differential equation for $u(r)$.

To carry out the numerical computation, we introduce the compactified radial
coordinate $x=1-2r_{H}/r$, which maps $r\in\lbrack r_{H},\infty)$ to
$x\in\lbrack-1,1]$. In this coordinate, the function $u(x)$ is approximated by
a truncated Chebyshev spectral expansion,
\begin{equation}
u\left(  x\right)  \simeq\sum_{n=0}^{N_{x}-1}c_{n}\,T_{n}(x),
\end{equation}
where $c_{n}$ are the spectral coefficients, $T_{n}(x)$ are the Chebyshev
polynomials of the first kind, and $N_{x}$ denotes the number of collocation
points, i.e., the spectral resolution. Evaluating the differential equation
for $u(x)$ at the Chebyshev collocation points yields a set of $N_{x}$ coupled
algebraic equations for the $N_{x}$ unknown coefficients $c_{n}$. To remove
the linear scaling invariance of Eq.~\eqref{eq:FDME}, we impose a
normalization condition by fixing $u(-1)=1$. This provides one additional
algebraic constraint. To maintain a square system, we treat the QNM frequency
$\omega$ as an additional unknown, so that the total number of equations
matches the total number of variables.

The resulting nonlinear algebraic system for $\{c_{n}\}$ and $\omega$ is
solved in the complex plane using the Newton-Raphson method. Starting from an
initial guess, each iteration linearizes the residual equations around the
current estimate and updates the solution by solving the associated linear
system for the correction. In our implementation, the linear system at each
step is solved with the built-in \texttt{LinearSolve} function in
\textsc{Mathematica}. The iteration is terminated when the difference between
successive iterates falls below $10^{-10}$, indicating numerical convergence.

\subsection{Waveform Extraction}

Alternatively, the QNMs of the scalar field $\Psi$ can be extracted from its
time-domain waveform. The waveform is obtained by numerically integrating
Eq.~\eqref{eq:TDME} with a second-order leapfrog method. Defining $\psi
_{i,j}\equiv\psi(i\Delta t,j\Delta r^{\ast})$ and $V_{j}\equiv V_{\mathrm{eff}%
}(j\Delta r^{\ast})$, the evolution is implemented via
\begin{equation}
\psi_{i+1,j}=-\psi_{i-1,j}+\frac{\Delta t^{2}}{\Delta r^{\ast2}}\left(
\psi_{i,j+1}+\psi_{i,j-1}-2\psi_{i,j}\right)  -\Delta t^{2}V_{j}\psi_{i,j}.
\end{equation}
We choose $\Delta t/\Delta r^{\ast}=1/2$ to ensure numerical stability. The
initial data are specified as a Gaussian pulse centered near the peak of the
effective potential,
\begin{equation}
\psi(t=0,r^{\ast})=\exp\!\left[  -\frac{(r^{\ast}-a)^{2}}{2\sigma^{2}}\right]
,
\end{equation}
where $a$ denotes the peak location, and we set $\psi(t<0,r^{\ast})=0$.

It is well known that the late-time waveform can be approximated as a
superposition of $N$ exponentially damped sinusoids associated with individual
QNMs,
\begin{equation}
\psi(t)\simeq\mathrm{Re}\sum_{n=0}^{N-1}A_{n}e^{-i\left(  \omega_{n}t-\phi
_{n}\right)  }. \label{eq:QNM}%
\end{equation}
Here $\omega_{n}$, $A_{n}$, and $\phi_{n}$ are the frequency, amplitude, and
phase of the $n$th mode, respectively. However, when two QNMs coalesce at an
EP, the QNM expansion is modified to \cite{PanossoMacedo:2025xnf}
\begin{equation}
\psi(t)\simeq\mathrm{Re}\sum_{n\neq\mathrm{EP}}A_{n}e^{-i\left(  \omega
_{n}t-\phi_{n}\right)  }+\mathrm{Re}\left[  \left(  A_{0}^{\mathrm{EP}%
}e^{i\phi_{0}^{\mathrm{EP}}}+A_{1}^{\mathrm{EP}}e^{i\phi_{1}^{\mathrm{EP}}%
}t\right)  e^{-i\omega_{\mathrm{EP}}t}\right]  , \label{eq:EP}%
\end{equation}
where $\omega_{\mathrm{EP}}$ denotes the QNM frequency at the EP. The
coefficients $A_{0}^{\mathrm{EP}}$ and $A_{1}^{\mathrm{EP}}$ characterize the
time-independent and linearly time-dependent contributions, respectively,
while $\phi_{0}^{\mathrm{EP}}$ and $\phi_{1}^{\mathrm{EP}}$ are the
corresponding phases. In the following, when extracting QNMs via waveform
fitting, we refer to the ansatzes in Eqs.~\eqref{eq:QNM} and \eqref{eq:EP} as
the standard and EP ansatzes, respectively. In addition, for the $n$th peak
(off-peak) mode, the corresponding frequency, amplitude, and phase are denoted
by $\omega_{n}^{p}$, $A_{n}^{p}$, and $\phi_{n}^{p}$ ($\omega_{n}^{o}$,
$A_{n}^{o}$, and $\phi_{n}^{o}$), respectively.

We employ two fitting schemes to extract QNMs from the time-domain waveform:
\emph{(i) Strong (agnostic) fit:} all parameters are treated as free, and we
use the \texttt{NonlinearModelFit} function in \textsc{Mathematica} to
determine them. The resulting QNM frequencies provide an independent
cross-check of the frequency-domain results. \emph{(ii) Weak fit:} the
frequencies are fixed to the values obtained in the frequency domain, while
only amplitudes and phases are treated as fitting parameters. In this case,
the fitting reduces to a linear regression problem \cite{Wang:2025mxe}, which
we solve using the \texttt{LinearModelFit} function in \textsc{Mathematica}.
The extracted amplitudes then quantify the relative contribution of each QNM
to the time-domain signal.

To assess the robustness of the extraction, we perform the fit over a sequence
of time windows with varying start times and a fixed end time. If the waveform
is well described by the chosen QNM ansatz, the fitted parameters exhibit a
stable plateau and remain nearly constant over a range of start times
\cite{Baibhav:2023clw,Takahashi:2023tkb}.

\section{QNM Spectrum}

\label{sec:QS}

The emergence of an EP in the QNM spectrum is generally tied to the variation
of at least two control parameters in the underlying non-Hermitian eigenvalue
problem. In the EMS model, this requirement is naturally fulfilled by the
coupling constant $\alpha$ and the BH charge $q$. Motivated by this structure,
in this section we carry out a frequency-domain search for EPs in the QNM
spectrum. Our strategy is to fix $\alpha$ and scan over $q$ to track the
evolution of neighboring QNM branches; if no coalescence is observed, we then
adjust $\alpha$ and repeat the scan. Following this procedure, for $l=2$ we
find a pair of QNMs that coalesce at an EP located at $\alpha=\alpha
_{\mathrm{EP}}\simeq0.79040751953125$ and $q=q_{\mathrm{EP}}\simeq
1.0421911480$.

\begin{figure}[ptb]
\includegraphics[width=1.03\linewidth]{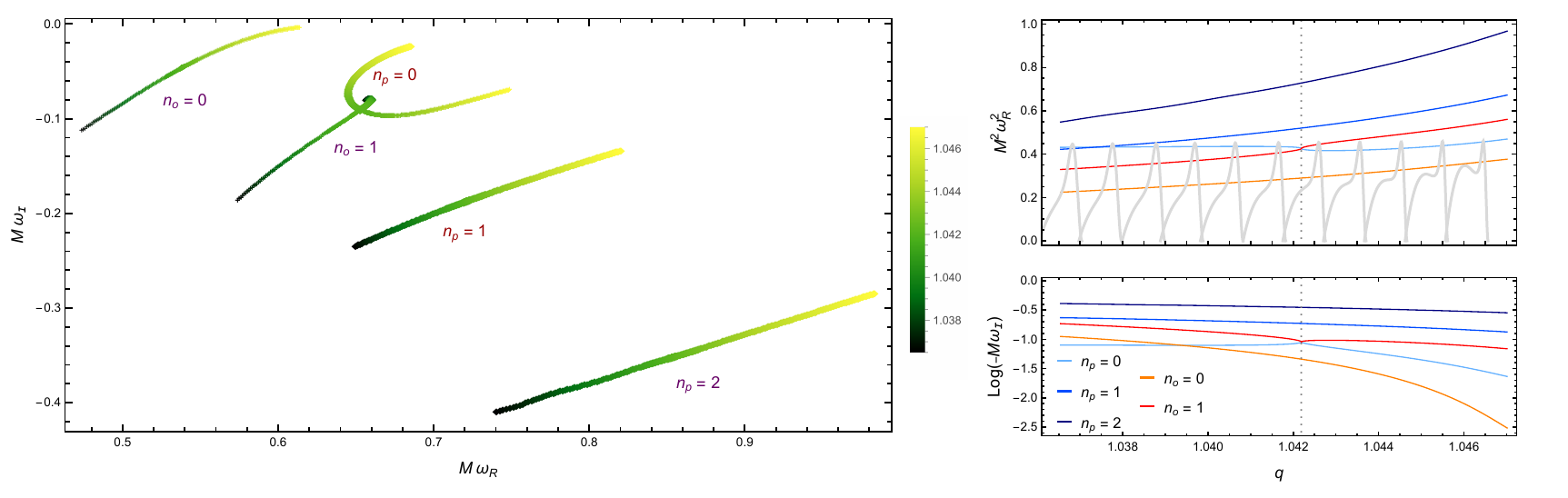}\caption{QNM spectrum of the
test scalar field at $\alpha=\alpha_{\mathrm{EP}}$ for $l=2$. \textbf{Left:}
Trajectories of QNM frequencies in the complex $\omega$ plane as the BH charge
$q$ is varied. The color bar indicates the value of $q$. The spectrum is
organized into five branches, two of which meet at an EP at $q=q_{\mathrm{EP}%
}$. Near the EP, each of the two participating branches splits into an upper
and a lower segment; the two upper (lower) segments recombine into the branch
labeled by $n_{p}=0$ ($n_{o}=1$). Peak and off-peak modes are indexed by
$n_{p}$ and $n_{o}$, respectively. \textbf{Upper right:} Squared real part of
the QNM frequencies, $\omega_{R}^{2}$, as a function of $q$. The gray dotted
line marks $q_{\mathrm{EP}}$, at which the real parts of the $n_{p}=0$ and
$n_{o}=1$ branches cross. The panel also shows the effective potential
$V_{\mathrm{eff}}$ for several representative values of $q$, with each curve
aligned so that its highest peak corresponds to the selected $q$. As $q$
increases, $V_{\mathrm{eff}}$ transitions from a single-peaked to a
double-peaked structure. In the single-peak regime, peak and off-peak modes
are distinguished by the location of $\omega_{R}^{2}$: the former lie close to
the peak value of $V_{\mathrm{eff}}$, whereas the latter take values
appreciably below it. \textbf{Lower right:} Imaginary part of the QNM
frequencies as a function of $q$. At $q=q_{\mathrm{EP}}$, the imaginary parts
of the $n_{p}=0$ and $n_{o}=1$ modes coincide.}%
\label{fig:qnmFD}%
\end{figure}

We display the QNM spectrum at $\alpha=\alpha_{\mathrm{EP}}$ in
Fig.~\ref{fig:qnmFD}. The left panel shows the trajectories of the QNM
frequencies in the complex $\omega$ plane as $q$ is varied. The color bar
indicates the value of $q$, with darker green corresponding to smaller $q$.
The spectrum is organized into five distinct branches, two of which meet at
the EP located at $q=q_{\mathrm{EP}}$. These QNM branches can be further
classified into peak and off-peak modes, labeled by $n_{p}$ and $n_{o}$,
respectively \cite{Wang:2025mxe}. A notable feature is that each of the two
branches involved in the EP is naturally divided by the EP itself into an
upper and a lower segment. The two upper segments recombine into the branch
labeled by $n_{p}=0$, whereas the two lower segments recombine into the branch
labeled by $n_{o}=1$. This branch reorganization reflects the local
square-root structure of the non-Hermitian spectrum near the EP and is
consistent with the mode switching or hysteresis behavior under parameter
variation around an EP \cite{Yang:2025dbn}. Hereafter, we refer to the two
modes involved in the spectral coalescence, namely the $n_{p}=0$ and $n_{o}=1$
modes, as the EP modes.

The upper-right and lower-right panels of Fig.~\ref{fig:qnmFD} show
$(M\omega_{R})^{2}$ and $\log(-M\omega_{I})$ as functions of $q$,
respectively, where $\omega_{R}\equiv\mathrm{Re}(\omega)$ and $\omega
_{I}\equiv\mathrm{Im}(\omega)$. The value of $q_{\mathrm{EP}}$ is marked by
the gray dotted line. The upper-right panel shows that the squared real parts
of the frequencies of the $n_{p}=0$ and $n_{o}=1$ branches cross at
$q=q_{\mathrm{EP}}$. It also displays the effective potential $V_{\mathrm{eff}%
}$ for several representative values of $q$. For sufficiently small $q$, the
effective potential has a single peak. The peak and off-peak modes labeled by
$n_{p}$ and $n_{o}$ correspond to modes localized near the potential peak and
appreciably below it, respectively.\footnote{\label{ft:1} The $n_{p}=2$ peak
mode would reside near the potential peak for values of $q$ smaller than those
shown in Fig.~\ref{fig:qnmFD}.}

The peak and off-peak modes are ordered by their decay rates, with $n_{p}=0$
and $n_{o}=0$ corresponding to the least damped peak and off-peak modes,
respectively. As shown in the lower-right panel, as $q$ increases, the
$n_{p}=0$ and $n_{o}=0$ modes undergo a fundamental-mode overtaking. Moreover,
the imaginary parts of the frequencies of the $n_{p}=0$ and $n_{o}=1$ modes
coincide at $q=q_{\mathrm{EP}}$. At $q=q_{\mathrm{EP}}$, the fundamental mode
is the $n_{o}=0$ mode, while the EP modes appear as the lowest overtone. The
lower-right panel also shows that branch reorganization occurs for the
$n_{p}=0$ and $n_{o}=1$ modes in the neighborhood of the EP.

\begin{figure}[ptb]
\includegraphics[width=0.48\linewidth]{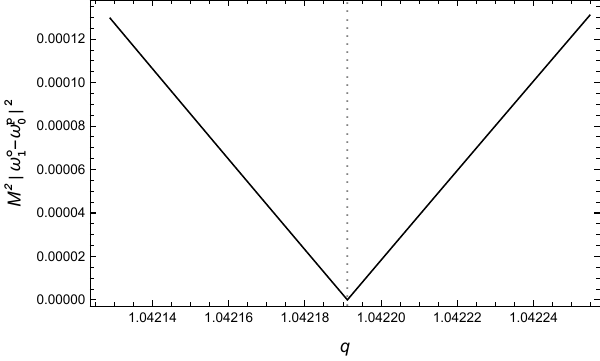} \hspace{6pt}
\includegraphics[width=0.48\linewidth]{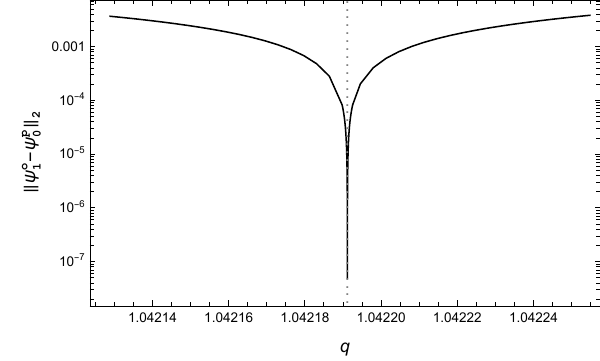}\caption{Squared difference
between the QNM frequencies $\omega_{0}^{p}$ and $\omega_{1}^{o}$ for the
$n_{p}=0$ and $n_{o}=1$ modes (\textbf{Left}), and the $L^{2}$ norm of the
difference between the corresponding eigenfunctions $\psi_{1}^{o}$ and
$\psi_{0}^{p}$ (\textbf{Right}), as functions of the BH charge $q$. In both
panels, the gray dotted line marks the EP location at $q=q_{\mathrm{EP}}$. As
$q$ approaches $q_{\mathrm{EP}}$, the squared frequency difference decreases
approximately linearly to an extremely small value, while the eigenfunction
difference drops sharply, indicating the near coalescence of the two
eigenfunctions. The simultaneous strong suppression of the eigenvalue and
eigenfunction differences provides direct numerical evidence for the EP in the
QNM spectrum.}%
\label{fig:EPeign}%
\end{figure}

To further resolve the QNM behavior in the immediate vicinity of the EP,
Fig.~\ref{fig:EPeign} shows the dependence on $q$ of both the frequency
splitting and the eigenfunction mismatch for the $n_{p}=0$ and $n_{o}=1$
modes. The left panel displays the squared frequency splitting, $\delta
\omega^{2}\equiv\left\vert \omega_{1}^{o}-\omega_{0}^{p}\right\vert ^{2}$, as
a function of $q$ near $q_{\mathrm{EP}}$, indicated by the gray dotted line.
We observe a pronounced cusp at $q=q_{\mathrm{EP}}$: $\delta\omega^{2}$
decreases approximately linearly as $q$ approaches $q_{\mathrm{EP}}$ from
either side and increases again after passing through the EP. This behavior is
characteristic of a second-order EP, where the local eigenvalue splitting
scales as the square root of the control parameter, so that the squared
splitting becomes linear sufficiently close to the EP \cite{Yang:2025dbn}. At
$q=q_{\mathrm{EP}}$, the numerical splitting does not vanish identically, but
remains below our numerical tolerance of $10^{-6}$.

The right panel shows the corresponding $L^{2}$ mismatch $\left\Vert \psi
_{1}^{o}-\psi_{0}^{p}\right\Vert _{2}$ between the two eigenfunctions
$\psi_{1}^{o}$ and $\psi_{0}^{p}$. As $q$ approaches $q_{\mathrm{EP}}$, this
quantity drops sharply to an extremely small value, indicating that the two
eigenfunctions become nearly indistinguishable. The simultaneous near
vanishing of the eigenvalue splitting and the eigenfunction mismatch provides
direct numerical evidence for the coalescence of both eigenvalues and
eigenfunctions, which is the defining signature of an EP in a non-Hermitian
QNM problem.

\section{QNM Extraction}

\label{sec:QE}

\begin{figure}[ptb]
\includegraphics[width=0.5\linewidth]{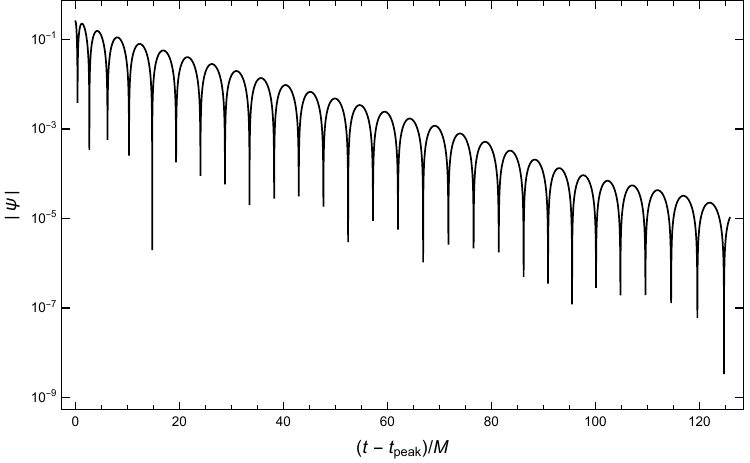}\caption{Time-domain ringdown
waveform $\psi$ of the scalar field at $\alpha=\alpha_{\mathrm{EP}}$ and
$q=q_{\mathrm{EP}}$, plotted as a function of $(t-t_{\mathrm{peak}})/M$. Here
$t_{\mathrm{peak}}$ denotes the time at which the waveform amplitude reaches
its maximum. The observer is located at $r=27.452M$.}%
\label{fig:3}%
\end{figure}

In this section, we extract QNMs from the time-domain waveform at
$\alpha=\alpha_{\mathrm{EP}}$ and $q=q_{\mathrm{EP}}$, where the
frequency-domain analysis indicates the presence of an EP associated with the
coalescence of two spectral branches. The waveform used in the extraction is
shown in Fig.~\ref{fig:3}. We consider both the standard and EP ansatzes,
given in Eqs.~\eqref{eq:QNM} and \eqref{eq:EP}, respectively, to assess how
the choice of ansatz affects the extracted frequencies and amplitudes. The
extraction is performed using both the weak fit and strong fit schemes. In
each case, we repeat the fit while varying the start time $t_{0}$ over the
interval $(t_{0}-t_{\mathrm{peak}})/M\in\lbrack0,130]$. Here $t_{\mathrm{peak}%
}$ denotes the time at which the waveform amplitude reaches its maximum.

\subsection{Weak Fit}

\begin{figure}[ptb]
\includegraphics[width=0.48\linewidth]{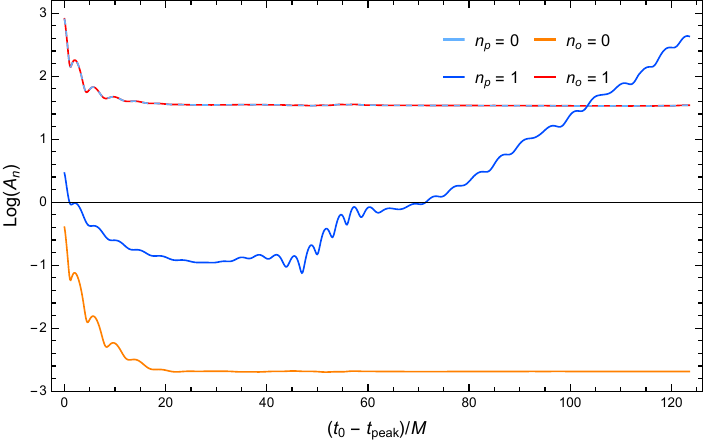} \hspace{6pt}
\includegraphics[width=0.48\linewidth]{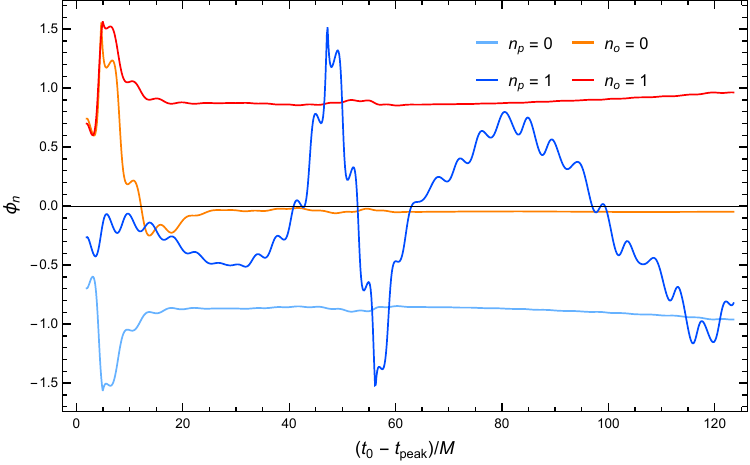}\caption{Extracted amplitudes
$A_{n}$ (\textbf{Left}) and phases $\phi_{n}$ (\textbf{Right}) obtained with
the standard ansatz, shown as functions of the start time $(t_{0}%
-t_{\mathrm{peak}})/M$. The fit includes the four least damped modes, namely
the $n_{p}=0,1$ peak modes and the $n_{o}=0,1$ off-peak modes, with
frequencies fixed by the frequency-domain analysis. The two EP modes,
$n_{p}=0$ and $n_{o}=1$, have nearly equal amplitudes and dominate over the
other fitted modes, while their phases are approximately opposite, indicating
destructive interference between their contributions.}%
\label{fig:4}%
\end{figure}

We first apply the weak fit scheme based on the standard ansatz in
Eq.~\eqref{eq:QNM} to extract the amplitudes and phases from the time-domain
waveform. In this fit, we include the four least damped modes, namely the
$n_{p}=0,1$ peak modes and the $n_{o}=0,1$ off-peak modes, with their
frequencies fixed by the frequency-domain analysis. The left and right panels
of Fig.~\ref{fig:4} show the extracted amplitudes $A_{n}$ and phases $\phi
_{n}$, respectively, as functions of the start time $(t_{0}-t_{\mathrm{peak}%
})/M$. For three of the four modes, namely $n_{p}=0$, $n_{o}=0$, and $n_{o}%
=1$, both the amplitudes and phases approach clear plateaus for $(t_{0}%
-t_{\mathrm{peak}})/M\gtrsim20$, indicating that these modes can be extracted
reliably within this time window. By contrast, the $n_{p}=1$ mode exhibits a
noticeably stronger dependence on $t_{0}$. Since it is the most strongly
damped of the four modes, its extraction is feasible only over a relatively
early interval, roughly $20\lesssim(t_{0}-t_{\mathrm{peak}})/M\lesssim40$, and
is therefore less robust.

A particularly noteworthy feature is that the EP modes, namely the $n_{p}=0$
and $n_{o}=1$ modes, have nearly equal extracted amplitudes, both exceeding
those of the other two modes by at least two orders of magnitude. This
hierarchy indicates that the contribution from each individual EP mode is
strongly enhanced in the ringdown signal. At the same time, these two dominant
modes have approximately opposite phases, so that their contributions nearly
cancel in the waveform. This pattern is consistent with recent analyses of
ringdown near an EP, in which the resonant pair is strongly amplified while
destructive interference between the dominant components suppresses the net
contribution to the waveform \cite{Oshita:2025ibu}.

\begin{figure}[ptb]
\includegraphics[width=0.48\linewidth]{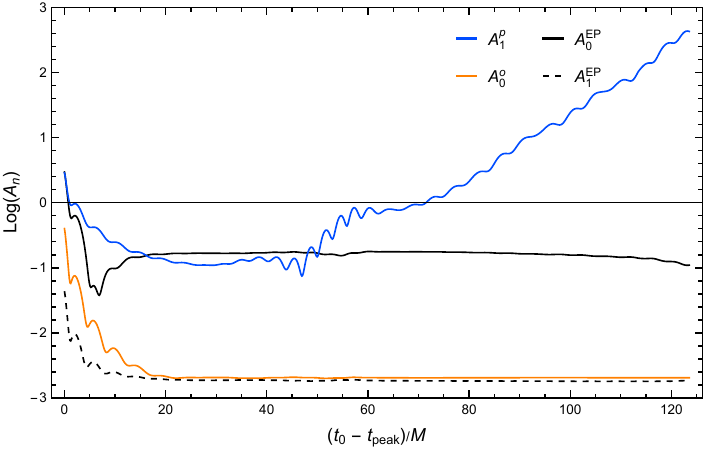} \hspace{6pt}
\includegraphics[width=0.48\linewidth]{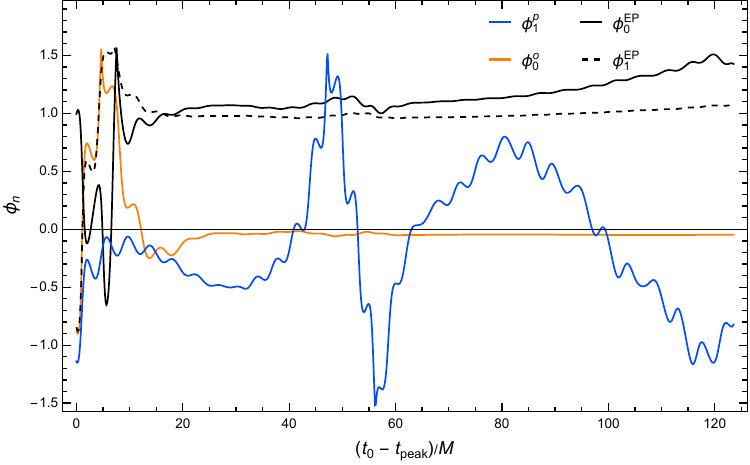}\caption{Extracted amplitudes
$A_{n}$ (\textbf{Left}) and phases $\phi_{n}$ (\textbf{Right}) obtained with
the EP ansatz, shown as functions of the start time $(t_{0}-t_{\mathrm{peak}%
})/M$. The fit includes two nonresonant modes, $n_{o}=0$ and $n_{p}=1$,
together with the EP contribution, with frequencies fixed by the
frequency-domain analysis. The EP contribution is decomposed into a
time-independent term (solid black) and a term linear in time (dashed black).
Their amplitudes are substantially smaller than the separate amplitudes of the
EP modes obtained with the standard ansatz in Fig.~\ref{fig:4}, indicating
that the EP ansatz provides a more economical description of the near-EP
ringdown.}%
\label{fig:5}%
\end{figure}

We next repeat the weak fit analysis using the EP ansatz, in which the two EP
modes are replaced by a single EP contribution, while the nonresonant modes
$n_{o}=0$ and $n_{p}=1$ are retained explicitly. The left and right panels of
Fig.~\ref{fig:5} show the extracted amplitudes and phases, respectively, as
functions of the start time $(t_{0}-t_{\mathrm{peak}})/M$. A notable feature
is that the EP contribution can be cleanly decomposed into a time-independent
term and a term linear in time, both of which exhibit relatively stable
plateaus over a broad fitting window. In contrast to the standard ansatz,
where the two EP modes are treated separately and acquire very large extracted
amplitudes, the EP ansatz absorbs their combined effect into a single resonant
contribution with substantially smaller coefficients. This behavior indicates
that the large amplitudes found in Fig.~\ref{fig:4} arise primarily from
forcing the near coalescing pair into two independent QNMs, whereas the EP
ansatz captures the ringdown more economically in terms of the resonant
structure near the EP. It is also worth noting that the time-independent and
time-linear EP terms have almost identical phases, so that they add nearly in
phase and together form a coherent resonant contribution. Altogether,
Fig.~\ref{fig:5} shows that the EP ansatz provides a more natural and stable
description of the waveform in the near EP regime. Moreover, since the
$n_{o}=0$ fundamental mode belongs to the off-peak family and is spectrally
less stable, its contribution is suppressed relative to the EP contribution
\cite{Wang:2025mxe}.

\subsection{Strong Fit}

In the strong fit scheme, we extract not only the amplitudes $A_{n}$ and
phases $\phi_{n}$, but also the frequencies $\omega_{n}$ of the fitted modes.
To assess the robustness of the extraction, we further introduce the frequency
deviation, $\delta\omega_{n}\equiv\left\vert \omega_{n}-\omega_{\mathrm{FD}%
}^{n}\right\vert $, which measures the difference between the frequency
extracted from the time-domain fit and the corresponding frequency-domain
value. In addition, for both the standard and the EP ansatzes, we consider
fits involving either one mode or two modes.

We first consider the standard ansatz in Eq.~\eqref{eq:QNM}. When only a
single QNM is included in the fit, the only mode that can be extracted
reliably is the one that asymptotically becomes the fundamental mode at late
times. In this regime, the overtones, including the EP modes, have already
decayed substantially, making their contributions to the waveform difficult to resolve.

\begin{figure}[ptb]
\includegraphics[width=1.01\linewidth]{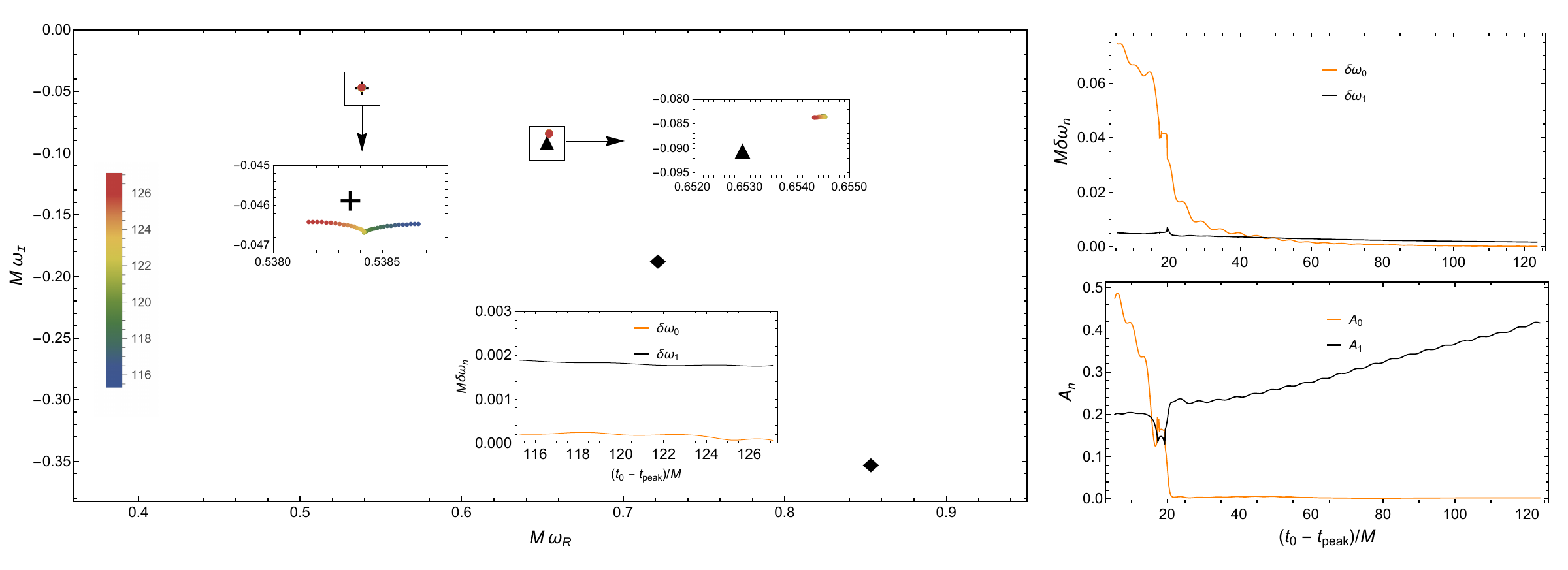}\caption{Results
of the strong fit using the standard ansatz with two modes included.
\textbf{Left:} Extracted QNM frequencies in the complex $\omega$ plane for
different start times, shown as shaded dots. The color bar indicates the start
time. The plus, triangle, and diamond markers denote the frequency-domain
values of the fundamental mode, the EP modes, and the peak-mode overtones,
respectively. The frequencies of the fundamental and EP modes can be extracted
reliably. \textbf{Upper right:} Frequency deviations $\delta\omega_{0}$ and
$\delta\omega_{1}$ for the fundamental and EP modes, plotted as functions of
$(t_{0}-t_{\mathrm{peak}})/M$. The inset in the left panel shows that their
minimum values are of order $10^{-4}$ and $10^{-3}$, respectively.
\textbf{Lower right:} Extracted amplitudes $A_{0}$ and $A_{1}$ for the
fundamental and EP modes, plotted as functions of $(t_{0}-t_{\mathrm{peak}%
})/M$. For $(t_{0}-t_{\mathrm{peak}})/M\gtrsim20$, $A_{0}$ remains
approximately constant, whereas $A_{1}$ increases nearly linearly with the
start time.}%
\label{fig:6}%
\end{figure}

The results of the two-mode fit with the standard ansatz are shown in
Fig.~\ref{fig:6}. They show that the frequencies of both the fundamental and
EP modes can be extracted reliably for $(t_{0}-t_{\mathrm{peak}})/M\gtrsim20$.
The left panel displays the extracted QNM frequencies in the complex $\omega$
plane for $115\lesssim(t_{0}-t_{\mathrm{peak}})/M\lesssim127$, where the
frequency deviations $\delta\omega$ are close to their minimum values. The
inset further shows that the frequency deviation $\delta\omega_{0}$ for the
fundamental mode reaches a minimum of order $10^{-4}$, while the corresponding
deviation $\delta\omega_{1}$ for the EP mode reaches a minimum of order
$10^{-3}$. More interestingly, the lower-right panel shows that, for
$(t_{0}-t_{\mathrm{peak}})/M\gtrsim20$, the extracted amplitude $A_{0}$ of the
fundamental mode remains approximately constant, whereas the extracted
amplitude $A_{1}$ of the EP mode increases almost linearly with the start
time. This behavior is precisely what one would expect from the term linear in
time appearing in the EP ansatz. In this sense, even when the waveform is
fitted with the standard ansatz, the resulting fit tends to reproduce the
characteristic structure encoded in the EP ansatz.

\begin{figure}[ptb]
\includegraphics[width=1\linewidth]{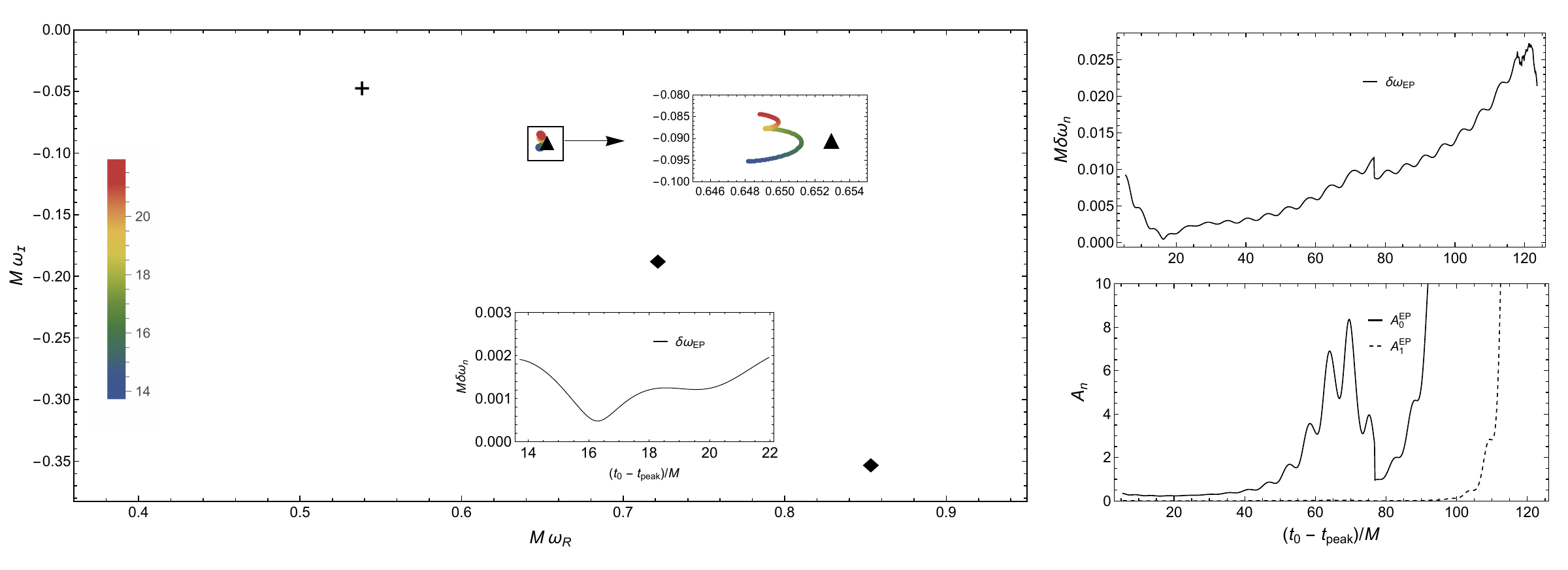}\caption{Results of the
strong fit obtained with the EP ansatz including only the EP terms.
\textbf{Left:} Extracted frequencies in the complex $\omega$ plane for
$14\lesssim(t_{0}-t_{\mathrm{peak}})/M\lesssim22$. Throughout this interval,
the extracted frequencies stay close to the frequency-domain value of
$\omega_{\mathrm{EP}}$. The inset shows that the frequency deviation
$\delta\omega_{\mathrm{EP}}$ attains a minimum of order $10^{-4}$ around
$(t_{0}-t_{\mathrm{peak}})/M\sim16$. The reference markers are the same as in
Fig.~\ref{fig:6}. \textbf{Upper right:} Frequency deviation $\delta
\omega_{\mathrm{EP}}$ as a function of the start time. \textbf{Lower right:}
Extracted amplitudes $A_{0}^{\mathrm{EP}}$ and $A_{1}^{\mathrm{EP}}$ as
functions of the start time. For $(t_{0}-t_{\mathrm{peak}})/M\lesssim40$, both
amplitudes remain approximately stable, indicating that the fit is reliable
within this time window. At later times, they become increasingly unstable,
accompanied by a growth in $\delta\omega_{\mathrm{EP}}$.}%
\label{fig:7}%
\end{figure}

Fig.~\ref{fig:7} shows the strong-fit results obtained with the EP ansatz
including only the EP contribution. The lower-right panel displays the
extracted amplitudes $A_{0}^{\mathrm{EP}}$ and $A_{1}^{\mathrm{EP}}$ as
functions of the start time. For $(t_{0}-t_{\mathrm{peak}})/M\lesssim40$, both
amplitudes remain approximately stable, indicating that within this time
window the waveform can be described reasonably well by the EP contribution
alone. At later times, however, the amplitudes become increasingly unstable,
while the frequency deviation $\delta\omega_{\mathrm{EP}}$ also grows. This
behavior suggests that the EP ansatz containing only the EP terms is effective
only over an early fitting window and becomes insufficient once the
fundamental mode begins to dominate the waveform. The left panel shows the
extracted frequencies in the complex $\omega$ plane as the start time varies
over $14\lesssim(t_{0}-t_{\mathrm{peak}})/M\lesssim22$, indicating that the
extracted frequencies remain close to the frequency-domain value of
$\omega_{\mathrm{EP}}$. The inset further shows that the frequency deviation
$\delta\omega_{\mathrm{EP}}$ reaches a minimum of order $10^{-4}$ around
$(t_{0}-t_{\mathrm{peak}})/M\sim16$.

\begin{figure}[ptb]
\includegraphics[width=1\linewidth]{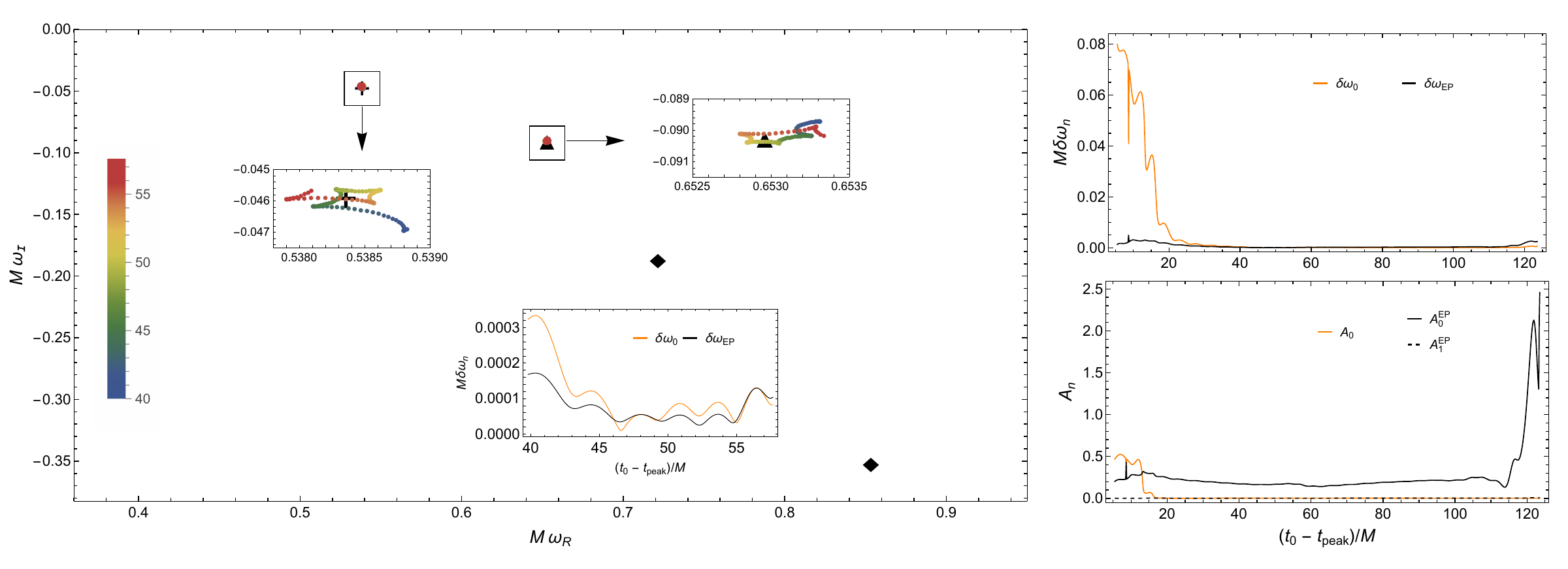}\caption{Results of
the strong fit obtained with the two-mode EP ansatz. \textbf{Left:} Extracted
frequencies in the complex $\omega$ plane as the start time varies over
$40\lesssim(t_{0}-t_{\mathrm{peak}})/M\lesssim60$, with the color bar
indicating $(t_{0}-t_{\mathrm{peak}})/M$. The reference markers are the same
as in Fig.~\ref{fig:6}. The extracted frequencies are concentrated near the
fundamental and EP modes. The inset shows that the minimum frequency
deviations are of order $10^{-5}$. \textbf{Upper right:} Frequency deviations
$\delta\omega_{0}$ and $\delta\omega_{\mathrm{EP}}$ as functions of the start
time. \textbf{Lower right:} Extracted amplitudes of the two contributions. The
EP contribution is shown in black, with $A_{0}^{\mathrm{EP}}$ plotted as a
solid line and $A_{1}^{\mathrm{EP}}$ as a dashed line. Compared with
Fig.~\ref{fig:7}, the inclusion of the additional mode leads to a more stable
extraction and yields broad plateaus for both contributions.}%
\label{fig:8}%
\end{figure}

We finally consider the two-mode strong fit based on the EP ansatz, in which
the waveform is described by the EP contribution together with one additional
QNM contribution. The results are shown in Fig.~\ref{fig:8}. The left panel
shows the extracted frequencies in the complex $\omega$ plane for
$40\lesssim(t_{0}-t_{\mathrm{peak}})/M\lesssim60$, demonstrating that the
extracted frequencies are concentrated near the fundamental and EP modes. The
inset shows that the minimum frequency deviations are of order $10^{-5}$.
Compared with the two-mode fit based on the standard ansatz, the minimum
frequency deviations of the fundamental and EP modes are smaller by about one
and two orders of magnitude, respectively. The lower-right panel shows the
extracted amplitudes of the two contributions as functions of the start time,
where the EP contribution is decomposed into $A_{0}^{\mathrm{EP}}$ and
$A_{1}^{\mathrm{EP}}$. In contrast to the single-mode fit in Fig.~\ref{fig:7},
the inclusion of the additional mode significantly improves the stability of
the extraction and yields broad plateaus for both contributions. This
indicates that, in this regime, the waveform is more accurately described by
the combined contribution of the fundamental and EP modes.

\section{Conclusion}

\label{sec:Con}

In this work, we investigated the QNM spectrum of a test massless scalar field
on hairy BHs in the EMS theory. By exploring the parameter space spanned by
the coupling constant $\alpha$ and the BH charge $q$, we identified an EP in
the frequency-domain spectrum at $\alpha=\alpha_{\mathrm{EP}}$ and
$q=q_{\mathrm{EP}}$, where two QNMs coalesce. Our analysis showed that this
coalescence occurs not only at the level of the complex frequencies, but also
at the level of the corresponding eigenfunctions. These results establish that
EPs can arise in the scalar QNM spectra of hairy BHs and provide a concrete
realization of non-Hermitian spectral degeneracies in BH perturbation theory.

We then investigated the time-domain signatures of the EP by extracting QNMs
from the ringdown waveform at $\alpha=\alpha_{\mathrm{EP}}$ and
$q=q_{\mathrm{EP}}$. Using both weak fit and strong fit schemes, we compared
the standard ansatz, based on the usual superposition of QNMs, with the EP
ansatz, which incorporates the resonant contribution associated with the
spectral degeneracy. In the weak fit analysis, with the frequencies fixed by
the frequency-domain results, we found that the amplitudes of the two
coalescing modes are much larger than those of the other modes, while
destructive interference between their contributions strongly suppresses their
net effect in the waveform. By contrast, the EP ansatz absorbs their combined
effect into a single resonant contribution consisting of a time-independent
term and a term linear in time, whose amplitudes are comparable to those of
the other modes.

In the strong fit analysis, where the QNM frequencies are treated as fitting
variables, the standard ansatz yields an amplitude that grows approximately
linearly with the start time, mirroring the linear time dependence encoded in
the EP ansatz. Moreover, the EP ansatz, especially in the two-mode fit, gives
significantly smaller differences between the extracted frequencies and their
frequency-domain counterparts. Together with the absence of the strong
amplitude hierarchy found in the standard ansatz, these results indicate that
the EP ansatz provides a more natural and robust description of ringdown
signals near EPs.

Overall, our results show that EPs in QNM spectra are not merely features of
the frequency-domain spectrum, but can also leave characteristic signatures in
ringdown signals. This highlights the importance of the non-Hermitian spectral
structure underlying BH perturbations and suggests that EP based descriptions
may be useful for analyzing ringdown in systems with nearly degenerate QNMs.
Possible extensions of the present work include the study of EPs in other
perturbation sectors or in other hairy BH models, as well as a more systematic
investigation of their implications for BH spectroscopy.

\begin{acknowledgments}
We are grateful to Yiqian Chen and Guangzhou Guo for helpful discussions and
valuable comments. This work was supported in part by NSFC (Grant Nos.
12275183, 12275184, and 12175212) and by Development Fund Project of Shanghai
University of Finance and Economics Zhejiang College for the Year 2024 (Grant
No. 2024FZJJ02).
\end{acknowledgments}

\bibliographystyle{unsrturl}
\bibliography{ref}

\end{document}